\documentclass[12pt]{article}
\usepackage{graphicx}

\newcommand\pubnumber{SNSN-323-63}
\newcommand\pubdate{\today}

\def\tulane{Department of Physics and Engineering Physics\\
Tulane University, New Orleans, LA 70118 USA}
\def\support{\footnote{Work supported by the U.S. National
Science Foundation grant PHY-1205266}}

\def\Title#1{\begin{center} {\Large #1 } \end{center}}
\def\Author#1{\begin{center}{ \sc #1} \end{center}}
\def\Address#1{\begin{center}{ \it #1} \end{center}}

\newcommand\pubblock{\rightline{\begin{tabular}{l} \pubnumber\\
         \pubdate  \end{tabular}}}
\newenvironment{Abstract}{\begin{quotation}  }{\end{quotation}}
\newenvironment{Presented}{\begin{quotation} \begin{center} 
             PRESENTED AT\end{center}\bigskip 
      \begin{center}\begin{large}}{\end{large}\end{center} \end{quotation}}

\def\beq{\begin{equation}}
\def\eeq#1{\label{#1}\end{equation}}
\def\eeqn{\end{equation}}

\def\beqa{\begin{eqnarray}}
\def\eeqa#1{\label{#1}\end{eqnarray}}
\def\eeqan{\end{eqnarray}}

\let\bar=\overbar

\def\Dslash{\not{\hbox{\kern-4pt $D$}}}
\def\dslash{\not{\hbox{\kern-2pt $\del$}}}

\def\msb{{\bar{\ssstyle M \kern -1pt S}}}

\begin{document}
\begin{titlepage}
\pubblock

\vfill
\Title{The Neutron Lifetime}
\vfill
\Author{ F.~E.~Wietfeldt\support}
\Address{\tulane}
\vfill
\begin{Abstract}
The decay of the free neutron into a proton, electron, and antineutrino is the prototype semileptonic
weak decay and the simplest example of nuclear beta decay. The nucleon vector and axial vector weak coupling
constants $G_V$ and $G_A$ determine the neutron lifetime as well as the strengths of
weak interaction processes involving free neutrons and protons that are important in astrophysics, cosmology,
solar physics and neutrino detection. In combination with a neutron decay angular correlation measurement, the
neutron lifetime can be used to determine the first element of the CKM matrix $V_{ud}$. Unfortunately the two
main experimental methods for measuring the neutron lifetime currently disagree by almost 4$\sigma$. I will present
a brief review of the status of the neutron lifetime and prospects for the future.
\end{Abstract}
\vfill
\begin{Presented}
The 8th International Workshop on the CKM Unitarity Triangle (CKM 2014)\\
Vienna, Austria  September 8--12, 2014
\end{Presented}
\vfill
\end{titlepage}
\def\thefootnote{\fnsymbol{footnote}}
\setcounter{footnote}{0}
A free neutron decays into a proton, electron, and antineutrino with a lifetime of about 880 s. This semileptonic weak
decay occurs because the neutron mass is slightly larger than that of the final state system. Because the mass difference and
hence the decay energy 1.29 MeV is so small, the details of this interaction at the quark level are unimportant and the process can be
effectively treated as a four-fermion interaction with the matrix element:
\begin{equation}
\label{E:betaHam}
{\cal M} = \left[ G_V \overline{p}\, \gamma_{\mu} n - G_A \overline{p}\, \gamma_5\gamma_{\mu} n\right]
\left[\overline{e}\, \gamma_{\mu}  \left( 1 +  \gamma_5 \right) \nu \right].
\end{equation}
The nucleon vector and axial vector effective weak coupling constants $G_V$ and $G_A$ determine the neutron decay rate and
therefore the neutron lifetime:
\begin{equation}
\label{E:tauN}
\tau_n =  \left(\frac{2 \pi^3 \hbar^7}{m_e^5 c^4 f_R}\right)\frac{1}{G_V^2 + 3 G_A^2} 
\end{equation}
where $f_R$ is a phase space factor that includes final state and radiative corrections. Conservation of vector current (CVC) requires that the vector weak coupling in the nucleon system has the same strength as for a bare quark, {\em i.e.}
$G_V = G_FV_{ud}$, where $G_F$ is the universal weak coupling constant obtained most precisely from the muon lifetime: 
$G_F = 1.1663787(6)\times 10^{-5}$ GeV$^{-2}$ \cite{PDG}, and $V_{ud}$ is the first element of the CKM matrix.
Axial current is not conserved so the value of $G_A$ is altered by the strong interaction
in the hadronic environment. Thus $G_A = G_FV_{ud}\lambda$, where $\lambda$ is measured experimentally from neutron decay. A measurement of the neutron lifetime $\tau_n$ along with $\lambda$ (via a neutron decay angular correlation measurement such as the beta asymmetry \cite{Abe14}) determines $G_A$, $G_V$, and using the known value of $G_F$, $V_{ud}$. This relationship, via Eq. \ref{E:tauN}, can be expressed in the following convenient form \cite{Mar06}:
\begin{equation}
 |V_{ud}|^2 = \frac{4908.7(1.9)s}{\tau_n (1 + 3\lambda^2)}
\end{equation}
Currently the most precise determinations of both $G_V$ and $V_{ud}$ come from the ${\cal F}t$ values of 13 superallowed $0^+\rightarrow0^+$ beta decay systems yielding $V_{ud} = 0.97425(22)$ \cite{Har09,Har13,Gar14}, a precision of $2\times 10^{-4}$, limited by theoretical uncertainties in the radiative, isospin breaking, and nuclear structure corrections. The neutron decay determination of $V_{ud}$ is, in principle, preferred as it is free of isospin breaking and nuclear structure corrections. The problem is the relatively worse precision and consistency in experimental results for $\lambda$ and $\tau_n$. 
\par
After the Big Bang, neutrons and protons were in thermal equilibrium via semileptonic weak interactions until the universe expanded to where the lepton density and temperature were too low to maintain equilibrium. This is called nucleon ``freeze out'', at about $t=1$ s. The ratio of neutrons to protons was then fixed by a Boltzmann factor: $n/p = \exp( -\Delta m / k T_{\rm freeze})\approx\frac{1}{6}$. The neutron lifetime directly provides the combination $G_V^2 + 3G_A^2$ that determines the semileptonic weak interaction rate and hence $T_{\rm freeze}$, the temperature of the universe at ``freeze out''. It also gives the fraction of neutrons that free decayed or were removed by lepton capture prior to the onset of light element nucleosynthesis. Nearly all neutrons that survived found themselves bound into $^4$He nuclei by $t=5$ min. The neutron lifetime experimental uncertainty is primarily responsible for the theoretical uncertainty in the primordial helium abundance $Y_P$ \cite{Bur99}.
\par
The most important neutron lifetime experiments to date have relied on two different methods: the beam method and the ultracold neutron (UCN) bottle method. A third approach, magnetic storage, has also been tried and will play an important role in future experiments. Further details on the history of neutron lifetime measurements with full discussion of these methods can be found in two recent review articles \cite{Dub11,Wie11}.
\par
\begin{figure}[htb]
\centering
\includegraphics[width=4.0in]{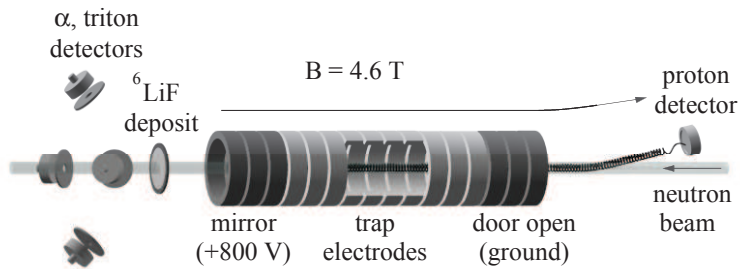}
\caption{An illustration of the Sussex-ILL-NIST beam neutron lifetime method.}
\label{F:tNScheme}
\end{figure}

The beam method is epitomized by the Sussex-ILL-NIST program which has produced the most recent and precise results for this method over the past thirty years \cite{Byr80,Byr90,Byr96,Dew03,Nic05,Yue13}. A schematic is shown in Fig, \ref{F:tNScheme}. A cold neutron beam passes through a quasi-Penning trap that consists of a 4.5 T axial magnetic field and a series of annular electrodes, three on each end held at +800 V with the central electrodes at ground. The maximum recoil energy of a neutron decay proton is 751 eV so any proton born in the central region is trapped. Periodically, typically every 20 ms, the three ``door'' electrodes are lowered to ground and a small ramped potential is applied to the central region to flush out trapped protons, which follow a 9.5$^{\circ}$ bend in the magnetic field and are accelerated and counted by a silicon surface barrier detector held at a large negative potential (-30 kV). The electrodes are then restored to the trapping state and the cycle is repeated. The neutron flux is measured by counting alphas and tritons from the $(n,\alpha)$ reaction in a deposit of $^6$LiF on a thin silicon crystal wafer. The proton count rate $R_p$ is given by the neutron decay rate in the effective trap volume and a proton detection efficiency factor $\varepsilon_p\approx 1$:
\begin{equation}
\label{E:Rp}
R_p = \varepsilon_p\frac{dN}{dt} = \varepsilon_p \frac{V_{\rm trap}}{\tau_n}\rho_n = 
\varepsilon_p \frac{A_{\rm beam} L_{\rm trap}} {\tau_n} \int \frac{\phi(v)}{v} dv
\end{equation}
where $V_{\rm trap}$ is the effective trap volume, the product of neutron beam area $A_{\rm beam}$ and effective trap length $L_{\rm trap}$; and $\rho_n$ is the neutron density, equal to the integral of spectral flux $\phi(v)$ over neutron velocity $v$. The neutron count rate $R_n$ is:
\begin{equation}
\label{E:Rn}
R_n = \varepsilon_{th} A_{\rm beam} v_{th} \int \frac{\phi(v)}{v} dv
\end{equation}
where $\varepsilon_{th}$, which contains the reference thermal cross section $\sigma_{th}$, is the efficiency for counting a neutron at the reference thermal velocity $v_{th} \equiv$ 2200 m/s. The spectral flux integral here comes from the $1/v$ law of neutron absorption. In the ratio $R_n/R_p$ these integrals cancel; the $1/v$ probability of neutron absorption precisely compensates for the $1/v$ probability for a neutron to decay in the trapping volume, so a high flux broad spectrum neutron beam (``white'' beam) can be used without need, in principle, to know the spectral flux (it is required for several small but important corrections however). Combining Eqs. \ref{E:Rp}, \ref{E:Rn} we have:
\begin{equation}
\tau_n = \frac{R_n \varepsilon_p L_{\rm trap}} {R_p \varepsilon_{th} v_{th}}.
\end{equation}
The neutron lifetime is obtained from two counting rates $R_p$, $R_n$ and two efficiencies $\varepsilon_p$, $\varepsilon_{th}$ all measured in the experiment. The effective trap length $L_{\rm trap}$ is complicated by end effects that can be removed by extrapolation -- it is for this reason that the trap is segmented and a range of different trap lengths are used.

\par
In the bottle method, ultracold neutrons (UCN, kinetic energy less than $\sim$200 neV) are stored in a suitably prepared material bottle. The effective neutron
potential energy in many materials is in the range 100--300 neV so UCN with less energy cannot penetrate and are completely reflected at the walls of the bottle, aside for a small but significant probability for upscattering or absorption. The general procedure is: 1) fill the bottle from a UCN source in a reproducible way; 2) store the UCN for a variable storage time $\Delta t$; 3) empty the bottle and count the remaining UCN, {\em e.g.} in a $^3$He proportional counter; 4) repeat steps 1--3 using different wall collision rates to account for wall losses (upscattering, absorption) by extrapolation. The wall collision rate can be modulated by either changing the surface/volume ratio of the bottle, such as with a piston; or by varying the initial UCN velocity spectrum, typically using gravity (for neutrons $mg$ = 103 neV/m). The most precise reported neutron lifetime measurement is the UCN bottle experiment of Serebrov, {\em et al.} \cite{Ser05} which used a bottle coated with a cryogenic fluoropolymer oil to obtain a significant reduction in wall losses relative to previous experiments, and employed a pair of storage bottles, one spherical and one cylindrical, to change the surface/volume ratio. 
\par
\begin{figure}[htb]
\centering
\includegraphics[width=4.8in]{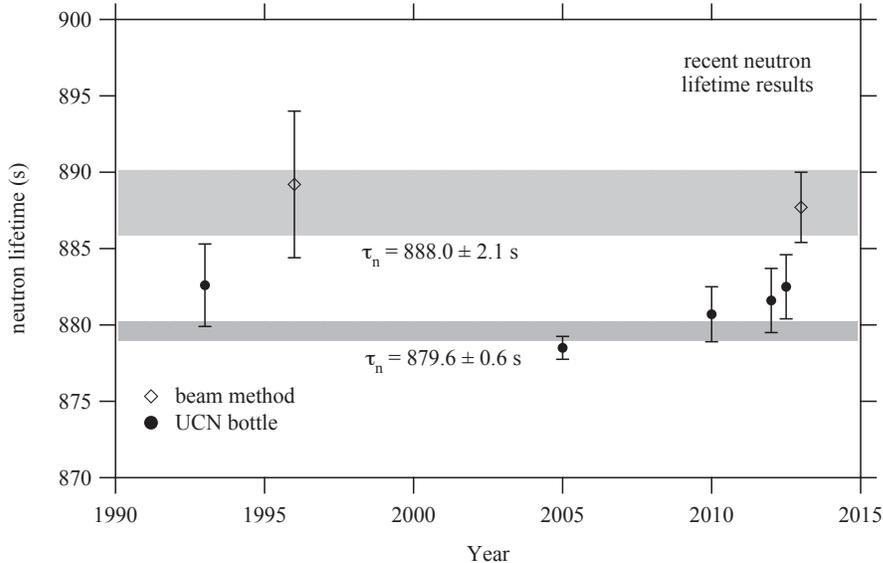}
\caption{A summary of recent neutron lifetime measurements, showing the five UCN bottle \cite{Mam93,Ser05,Pic10,Ste12,Arz12} and two neutron beam \cite{Byr96,Yue13} results used in the 2014 PDG recommended value of \mbox{$\tau_n = 880.3\pm1.1$ s}. The shaded regions show the weighted average $\pm1\sigma$ of each method, which disagree by 3.8$\sigma$.}
\label{F:nlifetimes}
\end{figure}

Fig. \ref{F:nlifetimes} shows a summary of recent neutron lifetime measurements, some of which are reevaluations and corrections to previously reported results. The 2014 Particle Data Group recommended value is $\tau_n = 880.3\pm1.1$ s \cite{PDG}, including an uncertainty scale factor of 1.9 to account for the overall lack of consistency. Considering the two methods separately we see that they agree among themselves, but their weighted averages disagree by 
$8.4 \pm 2.2$ s, a 3.8$\sigma$ discrepancy. Underestimated and/or unknown systematic effects in either or both methods are the most likely cause, although exotic physics explanations, such as mirror neutron oscillation \cite{Ber96}, have been proposed.
\par
Future neutron lifetime experiments will focus on testing for and studying systematic effects in both methods that could be responsible for the disagreement, while pushing the overall uncertainty down to the 0.1 s level. Next generation UCN bottle experiments are planned that use magnetic storage to avoid the loss mechanisms associated with material wall interactions. An improved and significantly larger (for higher statistics) beam method apparatus is being planned for the NIST program. If the neutron community can resolve these disagreements and reduce the relative uncertainties in the values of $\lambda$ and $\tau_n$ to the 10$^{-4}$ level then perhaps a value for $V_{ud}$ competitive in precision with $0^+\rightarrow0^+$ beta decay will be obtained that is theoretically much cleaner. Note that in this case both the beta decay and neutron decay results will be limited by the present theoretical uncertainty in the transition-independent radiative correction  $\Delta^V_R$ = .02361(38) \cite{Mar06}, which effects them both, so further theoretical work is also important.

\end{document}